\documentclass{elsart}
\usepackage{epsfig}
\begin{document}
\begin{frontmatter}
\title{High Level Tracker Triggers for CMS}
\author[Danek]{Danek Kotli\'nski}
\ead{danek.kotlinski@psi.ch}
\author[Andrei]{Andrey Starodumov\thanksref{itep}}
\ead{andrey.starodumov@pi.infn.it}
\address[Danek]{Paul Scherrer Institut, CH-5232 Villigen, Switzerland}
\address[Andrei]{INFN Sezione di Pisa, Via Livornese 1291, 
 56010 S.Piero a Grado(PI), Italy}
\thanks[itep]{On leave from ITEP, Moscow, Russia}
\begin{abstract}
Two fast trigger algorithms based on 3 innermost hits in the CMS 
Inner Tracker are presented. 
One of the algorithms will be applied at
LHC low luminosity to select B decay channels. 
Performance of the algorithm is demonstrated 
for the decay channel $B_{s}^0 \rightarrow D_{s}^-+\pi^+$.
The second algorithm will be used to select $\tau$-jets at LHC 
high luminosity. 
\end{abstract}
\begin{keyword}
Triggering \sep B-physics \sep Pixel detectors \sep CMS
\PACS 29.40.Gx 
\end{keyword}
\end{frontmatter}
%


\section{Introduction}
\label{int}
The collision rate at LHC \cite{lhc} is foreseen to be 40MHz. To handle such 
rates a powerful DAQ system and also fast and efficient 
trigger algorithms are needed. 
The CMS Collaboration \cite{cms} has chosen 
a trigger scheme based on two components: a hardware Level 1 Trigger
and a software Level 2 Trigger. As any collider detector, CMS is composed of
tracker, calorimeter and muon systems.  At Level 1 the information from the 
muon and calorimeter systems is used. 
At Level 2, called Higher Level Trigger (HLT) in CMS, 
also the tracker contributes to the event selection. 

The CMS tracker system \cite{TTDR} is based on two technologies: 
1) silicon pixel detectors, which are closest to the beam, provide 3-D hits
used for vertexing, and 2)
silicon strip detectors placed at larger radii are used for 
pattern recognition and track reconstruction.
The amount of information coming from the pixel detector is about 10\% of 
the full tracker, hence it is reasonable to use it at
the beginning of the HLT algorithms. 
In this article we present track reconstruction algorithms based on 3 
innermost tracker hits. At low LHC luminosity 
($10^{33}\mathrm{cm^{-2} s^{-1}}$), when only two 
pixel layers are foreseen, the third hit is taken from the lowest radius
strip detector layer. At high luminosity 
($10^{34}\mathrm{cm^{-2} s^{-1}}$) the algorithm is based purely on the  
pixel detector.


\section{Detector layout}
\label{det}

The results shown here were obtained with a full GEANT based simulation of 
the CMS detector using the CMSIM \cite{cmsim} and ORCA \cite{orca} software 
packages.
The CMS inner tracker layout used in the simulation will be briefly mentioned 
below, more details can be found in Ref.\cite{TTDR,Lenzi,Angarano,Steve}.
The pixel detector consists of
three barrel layers located at mean radii 4.3~cm, 7.2~cm and 11.0~cm.
The 52~cm long pixel barrel is supplemented by two endcap disks on each side.
With this configuration the pixel detector
provides 3 hit coverage up to rapidity $|\eta|\leq2.1$.
For low luminosity only two barrel layers will be installed
giving 2 pixel hits up to $|\eta|\leq2.1$.
The $1^{st}$  silicon strip barrel layer used in our reconstruction 
algorithm is 89.2cm long and is located at a mean radius of 23.5cm 

\section{The triggering scheme}

\label{daq}
The goal of the CMS trigger system is to reduce the event rate from 
$\mathrm{40MHz}$ down to $\mathrm{100Hz}$ that will be stored for 
off-line analysis. The DAQ architecture is designed in such 
a way that it should sustain maximum output rate of Level 1 trigger of 
$\mathrm{100kHz}$ and provide HLT acceptance of about $10^{-3}$. 
As mentioned in the Introduction the Level 1 trigger decision is based on 
the muon and calorimeter systems, at HLT the tracker information will
be added.
Even if the full detector information is accessible at HTL, fast algorithms 
which select event topologies of interest and use partial information
should be applied first. 
    
\section{Track and vertex reconstruction}
\label{trk}
Two algorithms based on the track reconstruction from the 3 innermost 
tracker layers will be discussed below,
both start from the two innermost pixel layers. 
For the $3^{rd}$ hit the first strip layer is used in one algorithm,
the second algorithm uses the third pixel layer.

\subsection{Track reconstruction using the pixel detector}
\label{pix}

\begin{figure}[hbtp]
  \begin{center}
    \resizebox{8cm}{!}{\rotatebox{-90}{\includegraphics{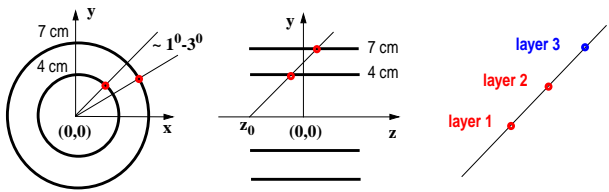}}}
    \caption{ Pixel detector track finding algorithm}
    \label{fig:pixel_algo}
  \end{center}
\end{figure}

The pixel track finding algorithm has been explained in
Ref.\cite{pix_reco}, here only the most important details will be given.
It's three essential steps are shown in Fig.~\ref{fig:pixel_algo}.
Pixel hit pairs from the first two layers (barrel+barrel or barrel+endcap)
are matched in $r-\phi$ and $z-r$ to establish track candidates.
The cuts are optimized for a minimum track transverse momentum ($p_{t}$) 
of $1GeV/c$ and permit a maximum impact parameter in $r-\phi$ of $1mm$. 
In the $z-r$ plane the hit pairs have to point to a region in $z$ within 
$3\sigma (+/-15cm)$ of the
LHC luminous region.                                     
Valid pixel pairs are matched with a $3^{rd}$ pixel hit forming a track 
candidate.
Using these tracks a list of primary vertices (PV) is formed at z values
where at least 3 tracks cross the z axis.
Tracks which do not point to any PV candidate are erased.
Due to the detector overlaps in the $r-\phi$ direction for some Monte Carlo
tracks more than one track candidate is found.
Track pairs which share pixel hits and are closer than
$10mrad$ from each other are identified.
A cleaning procedure then erases one of the tracks in each pair.
\begin{figure}[hbtp]
  \begin{center}
    \resizebox{8cm}{!}{\includegraphics{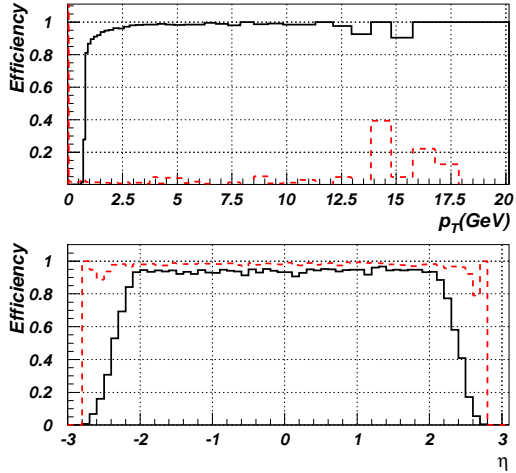}}
        \caption[]{The track reconstruction efficiency at high LHC luminosity 
        shown as a function of track $p_{t}$ (upper part) and rapidity 
        (lower part).
        The dashed line in the upper part shows the ghost track rate.
        The dashed line in the lower part shows the algorithmic efficiency
        defined as the ratio of the correctly reconstructed tracks to all 
        MC tracks with at least 3 pixel hits (see text for more details). 
        }
    \label{fig:track_eff}
  \end{center}
\end{figure}       
The track finding efficiency is illustrated 
in Fig.~\ref{fig:track_eff} where the efficiency (the ratio of reconstructed 
to Monte Carlo tracks) is shown as a function of the track $p_{t}$ and 
rapidity.
Only reconstructed tracks which have all 3 hits correctly assigned are
counted and the fraction is normalized to all Monte Carlo tracks which
are within the full acceptance of the pixel detector.
The plotted efficiency is for high luminosity LHC events in the presence
of minimum bias pile-up interactions (on the average 20).
However, the comparison is done for the signal tracks only, that is tracks 
originating from the minimum-bias events are ignored.
In the upper part of Fig.~\ref{fig:track_eff} the reconstruction efficiency
and the ghost rate is shown as a function of $p_{t}$.
The lower part shows that the efficiency (for tracks with $p_{t}\geq1GeV/c$) 
is flat within the rapidity coverage of the pixel detector. The solid curve 
shows the absolute efficiency and the dashed curve the algorithmic efficiency
(normalized to the number of tracks with 3 pixel hits). 
For all event types considered in our HTL studies the absolute track finding
efficiency for tracks above $1~GeV/c$ is about 90\%.
The algorithmic efficiency is between 93\% and 95\% and
the fraction of ghost tracks is typical between 5\% and 8\%.

Obviously the track parameters, fitted with 3 points only,  are much inferior
to the full tracker resolution. The reconstructed track's direction
is good, the resolution in the $\phi$ direction being $8mrad$
and in the $\theta$ direction $11mrad$.
The $p_{t}$ measurement however, is poor, mainly because of the small
radius of the last hit (11~cm). For $2GeV/c$ tracks this resolution  is about
7\%, but for $10GeV/c$ it reaches 20\%
(see Ref.\cite{pix_reco} for more details).
The curvature of the tracks is sufficient to distinguish the
track sign with 100\% efficiency up to $p_{t}$ of $30\div40~GeV/c$.  
\begin{figure}[h]
\centering
   {\mbox{ \epsfig{file=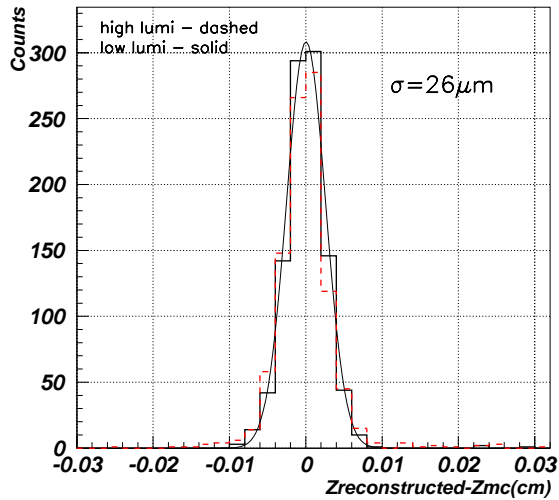,width=8.cm}}}
          \caption{The difference (in cm) between the z position of the Monte 
Carlo primary vertex and the reconstructed primary vertex. 
Only pixel hits are used in the vertex reconstruction.} 
\label{fig:vtx}
\end{figure} 
The primary vertex candidates found during the track finding stage are
reanalyzed. Only PVs with at least 3 valid tracks are kept and
the position of each vertex is estimated as the mean value of the z
impact parameters of all tracks assigned to it.
In addition to the main ``signal'' PV, on the average $6\div8$ 
 more PVs per event are found at high luminosity.
The ``signal'' PV is usually found with a high accuracy of better 
than 50~$\mu$m.
This is shown in Fig.~\ref{fig:vtx} where the difference in the
z position of the reconstructed vertex and the Monte Carlo vertex is plotted.
The efficiency of this algorithm is high, for the HLT event samples 
the ``signal'' PV is always found with an efficiency of better than 95\%.
                                                                             
\subsection{Reconstruction using the pixel detector and one strip layer}
\label{str}
\begin{figure}
\begin{center}
\includegraphics*[width=6cm]{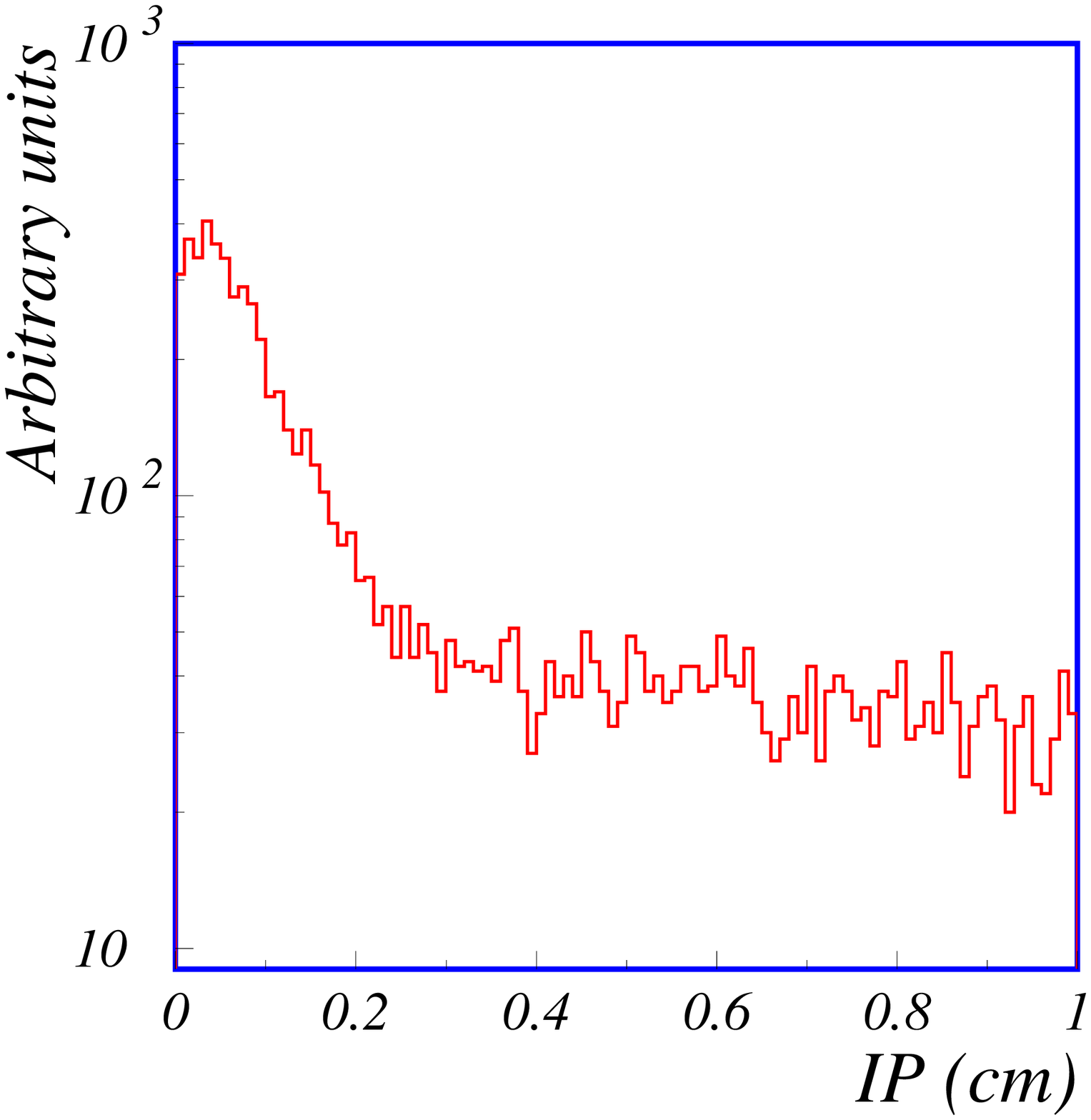}
\includegraphics*[width=6cm]{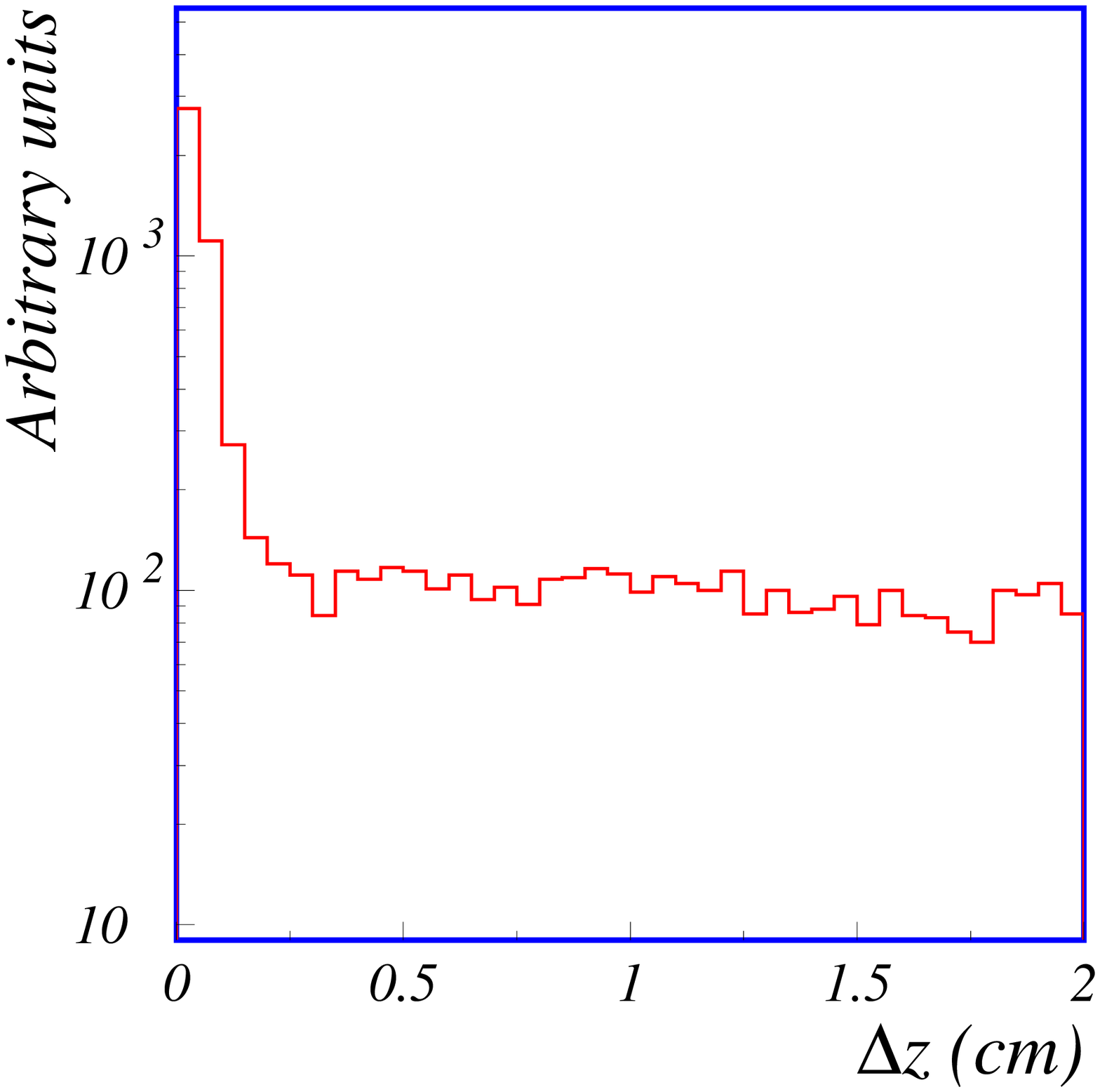}
\end{center}
\caption{Impact parameter (left) and
$\Delta z$ (right) distributions for final state tracks of decay (\ref{eq:xs}).}
\label{fig:ipz}
\end{figure}

This algorithm is similar to the one discussed in  
section~\ref{pix}. Since the algorithm is developed to trigger 
on exclusive B meson decays,
selection criteria listed below are optimized for such decays.
Below we describe it in more detail.   
\begin{enumerate}
\item Any pair of hits from two separate pixel layers are connected
with a straight line in the transverse plane. We utilize the fact
that the tracks we want to reconstruct are relatively stiff ($p_{\mathrm{T}} 
\ge 0.9 \mathrm{GeV}/c$) 
and the lever arm is small ($\sim 7 \mathrm{cm}$). 
Then the impact parameter (IP, the distance from the line to the primary 
interaction point in the transverse plane) is calculated. 
If it is less than $\mathrm{2mm}$, the two hits create a track 
candidate.
\item  Extrapolating the straight line that connects two selected hits 
to the third (strip) layer in the $r-z$ plane, the expected 
position ($z_{\mathrm{exp}}$) of a third hit is calculated. 
If there is a reconstructed hit with its
coordinate $z_{\mathrm{rec}}$ close to $z_{\mathrm{exp}}$, 
$\Delta z=|z_{\mathrm{exp}}-z_{\mathrm{rec}}| \le 1.5\mathrm{mm}$, 
it is taken as the third hit of the track candidate. The momentum
and the charge of the track is determined.
\end{enumerate}
In Fig.~\ref{fig:ipz} we show the $\mathrm{IP}$ and
$dz$  distributions for decay (\ref{eq:xs}) to illustrate the 
above cuts.
If the transverse momentum of the reconstructed track is greater then
 $0.9 \mathrm{GeV}/c$ it is accepted for further analysis.


\section{B-physics triggers}
\label{bph}

The Level 1 B physics trigger in CMS is mainly based on the 
muon system (the single muon trigger rate is assumed to be 10kHz). 
At the Level 2 information from the muon and 
calorimetric systems will be reanalyzed with fine granularity
and more sophisticated algorithms. 
However, soft products of B decays usually do not manifest themselves 
in the calorimeters and  more precise calculation of the trigger muon 
momentum can suppress the rate by only a factor of 2.
Thus, one needs to apply further selection algorithms which are specific
to B decays, e.g. reconstruction of B decay products (tracks)
and calculation of invariant masses of intermediate resonances if any. 

\begin{figure}
\begin{center}
\includegraphics*[width=7cm]{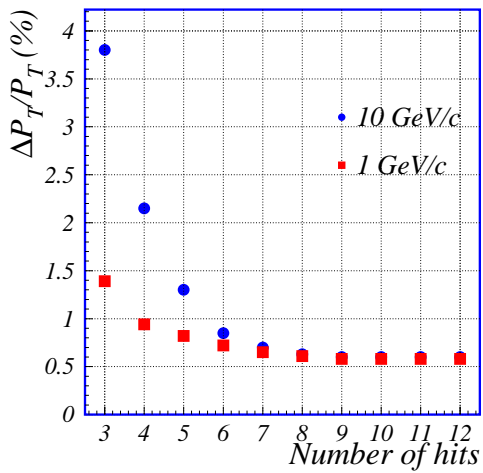}
\end{center}
\caption{Transverse momentum resolution as a function of number of hits for 
tracks with transverse momentum of 1 and 10GeV/c.}
\label{fig:pt3}
\end{figure}

The idea of using only 3 innermost hits at HLT to select the exclusive 
B decay modes is based on the observation that the momentum resolution 
provided by these 3 hits is only a few times worse than that obtained with 
the full tracker. 
Fig.~\ref{fig:pt3} shows the transverse momentum resolution in percentage
versus the number of continues hits used for track reconstruction.
Two distributions represent results for tracks of 1 and 10$\mathrm{GeV/c}$.
One can see that for 
tracks with $p_{\mathrm{T}} \simeq 1 \mathrm{GeV}/c$
the transverse momentum resolution obtained with 3 hits
is about 1.5\%, 2 times worse than with 
the full tracker.

The decay mode (and its charge conjugate)
\begin{equation}
 B_s^0 \rightarrow D_s^- + \pi^+ \rightarrow \phi + \pi^- + \pi^+
\rightarrow K^+ + K^- + \pi^- + \pi^+ 
\label{eq:xs}
\end{equation}
has been chosen as a benchmark. Only the $D_s$ decay chain is triggered 
at this stage which allows us to select also the following decay modes of 
the B meson:
\begin{equation}
B_s^0 \rightarrow D_s^- + X   
\label{eq:mu}
\end{equation}
where $X$ = $a_1^+$, $K^+$, $D_s^+$ and $\mu^+ +X$.

\begin{figure}
\begin{center}
\includegraphics*[width=6cm]{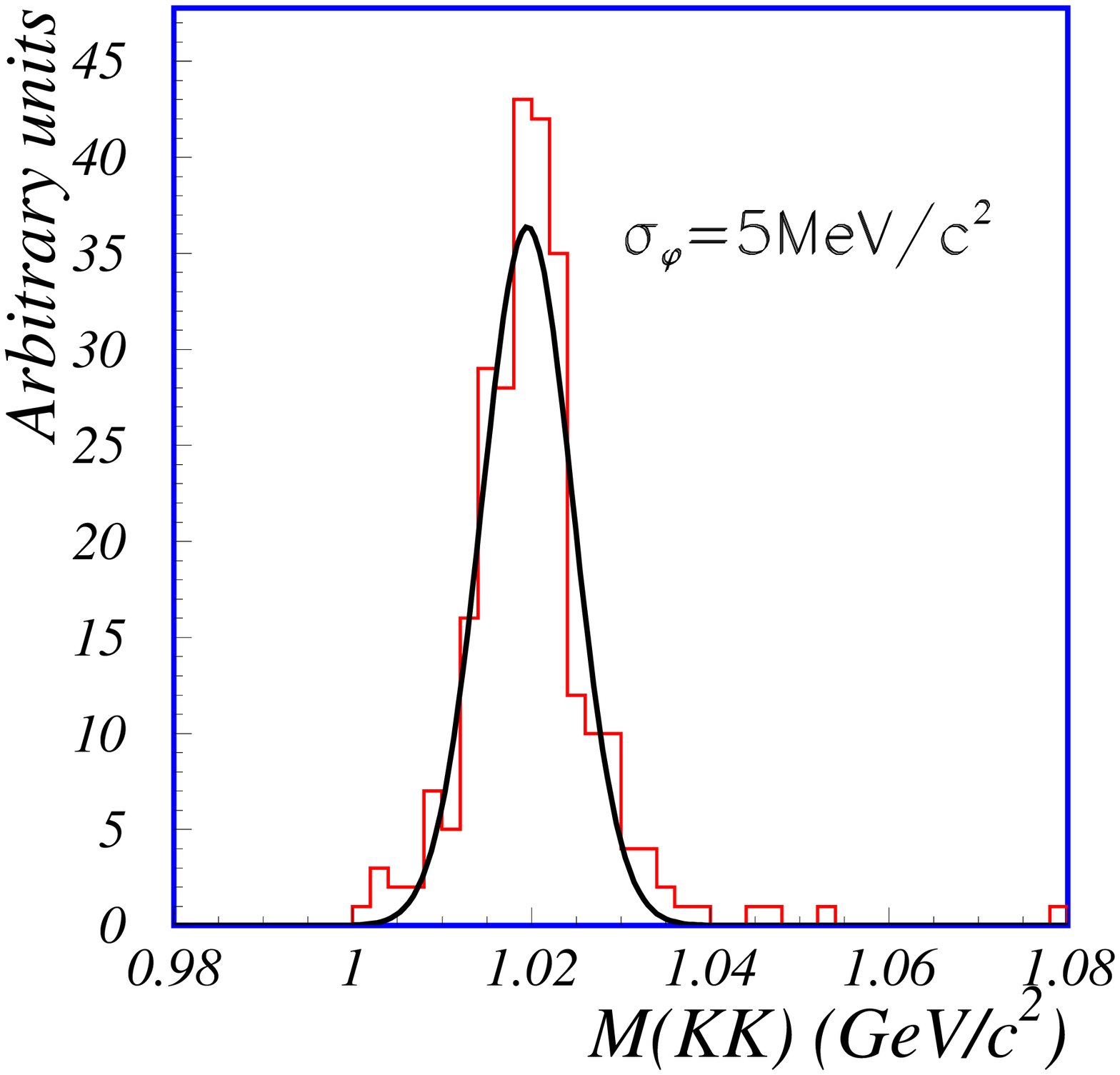}
\includegraphics*[width=6cm]{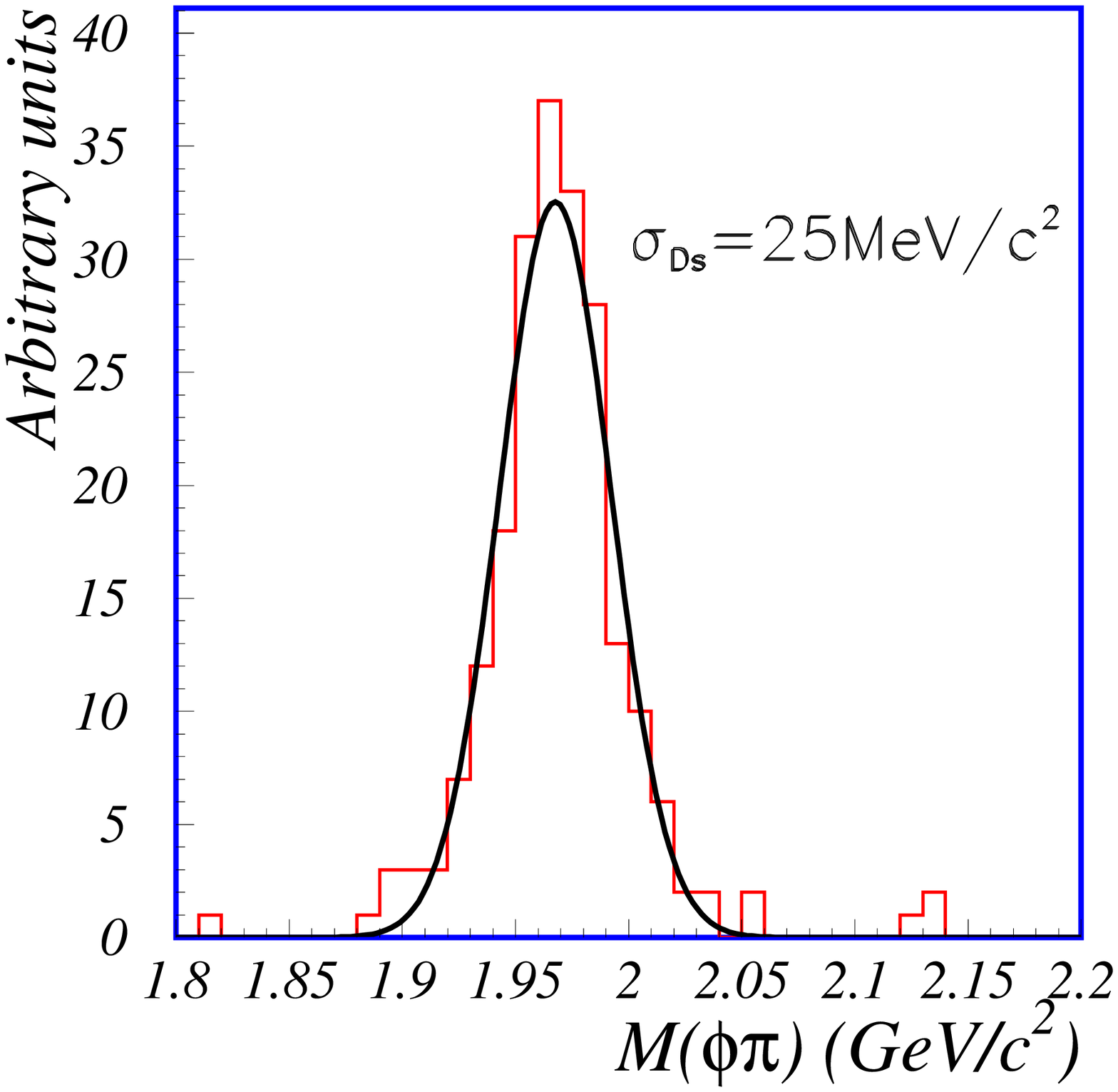}
\end{center}
\caption{Left - an invariant mass of two opposite charged tracks 
(assuming they are Kaons) and right -  an invariant mass of reconstructed
$\phi$ and adjacent $\pi$
 for the events of decay mode (\ref{eq:xs}).}
\label{fig:mass}
\end{figure}

Applying the reconstruction algorithm discussed in Section 
\ref{str} we obtain the momentum and the charge for each reconstructed track.
To trigger the decay modes (\ref{eq:xs},\ref{eq:mu})
we reconstruct all tracks with $p_t \ge 0.9GeV/c$ and then find all
$\phi$ and $D_s$ candidates by trying different track combinations.
An event is selected if two conditions are fulfilled. 
Firstly, that there are two oppositely charged tracks
close to each other ($\Delta R\le 0.3$), and their invariant mass 
(assuming they are kaons) is
compatible with the $\phi$ mass, $|M_{KK}-M_{\phi}| 
\le 15 \mathrm{MeV}/c^2$.
Secondly, there is a third track in the cone $\Delta R=1.0$
around the reconstructed $\phi$ candidate
and the invariant mass of this $\phi$ candidate and the third track
(assuming it is a pion) agrees with the
$D_s$ mass, $|M_{\phi\pi}-M_{D_s}| \le 70 \mathrm{MeV}/c^2$.
Mass cuts  have been defined based on expected
mass resolutions shown in Fig.~\ref{fig:mass} 
for the decay mode (\ref{eq:xs}): $\sigma(M_{\phi})
\simeq 5 \mathrm{MeV}/c^2$ and $\sigma(M_{D_s})\simeq 
25 \mathrm{MeV}/c^2$. 

The proposed algorithm 
reduces the single muon trigger rate  by factor of 30 while 
providing a signal efficiency of about 75\%. The fraction of ghost
tracks is relatively small (about 10\%). 
Further reduction of trigger rate will be based on more strict cuts
using the full CMS tracker capability to reconstruct exclusive 
B decays.

\section{Tau trigger}
\label{tau}
 
The ``pixel'' $\tau$ trigger proposed here is an example of a tracker 
HLT application which uses only the pixel data.
As signal events we used SUSY Higgs bosons decaying
into two $\tau$ leptons with two $\tau$ hadronic jet(s) in the final state. 
The main background for such events are QCD 2-jet events in the $p_{t}$ range
50-230~GeV/c.
\begin{figure}[hbtp]
  \begin{center}
    \resizebox{5cm}{!}{\rotatebox{90}{\includegraphics{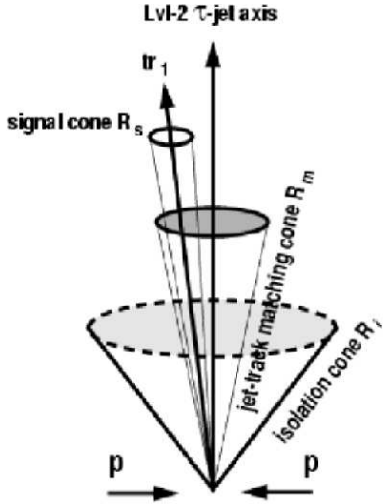}}}
        \caption[]{A sketch showing the basic principle of the pixel
        $\tau$-jet identification algorithm.}
    \label{fig:three_cones}
  \end{center}
\end{figure}     
More details about the algorithm and the event samples used in the simulation 
can be found in Ref.\cite{tau_trig}.\\
The HLT pixel algorithm is based on track isolation criteria and is
schematically shown in Fig~\ref{fig:three_cones}. 
The $\tau$-jet direction is defined by the calorimeter trigger.
All track candidates in the matching cone $R_{m}$ around the jet direction and
above the $p^{m}_{t}$ cut are considered in the search for signal tracks,
that is tracks which originate from the hadronic $\tau$ decay.
The track with the highest $p_{t}$ is declared the ``leading'' track
($tr_{1}$ in Fig.~\ref{fig:three_cones}).
Any other track which is in the narrow cone
$R_{s}$ around $tr_{1}$ is also assumed to come from the $\tau$ decay.
A larger area $R_{i}$ is now searched for tracks above the $p^{i}_{t}$ cut.
If no tracks are found in the $R_{i}$ cone, except the ones which are already
in the $R_{s}$ cone, the isolation criteria is fulfilled and the jet is
labeled as a $\tau$-jet.                           
The narrow signal cone $R_{s}$ around the ``leading'' track $tr_{1}$ is needed
in order to trigger on 3 prong $\tau$ decays in addition to 1 prong.
Typical values of the cuts used above are : $R_{s}$=0.05, $R_{m}$=0.10,
$R_{i}$=0.35, $p^{s}_{t}=3 GeV/$c and $p^{i}_{t}=1 GeV/c$.
                                                          
The algorithm works very well at low luminosity.
At high luminosity, however, it's efficiency becomes small (about 50\%)
due to the large number of tracks originating from the pile-up interactions.
The efficiency is improved by using the PVs information.
The vertex of the leading track is assumed to be the ``signal'' PV
(this assumption is correct for 99\% of events).
Once the ``signal'' PV is defined only those tracks which were
assigned to it are considered in the isolation criteria.
This approach increases the efficiency of our
algorithm at high luminosity to the same value as at low luminosity.
\begin{figure}[hbtp]
  \begin{center}
    \resizebox{10cm}{!}{\includegraphics{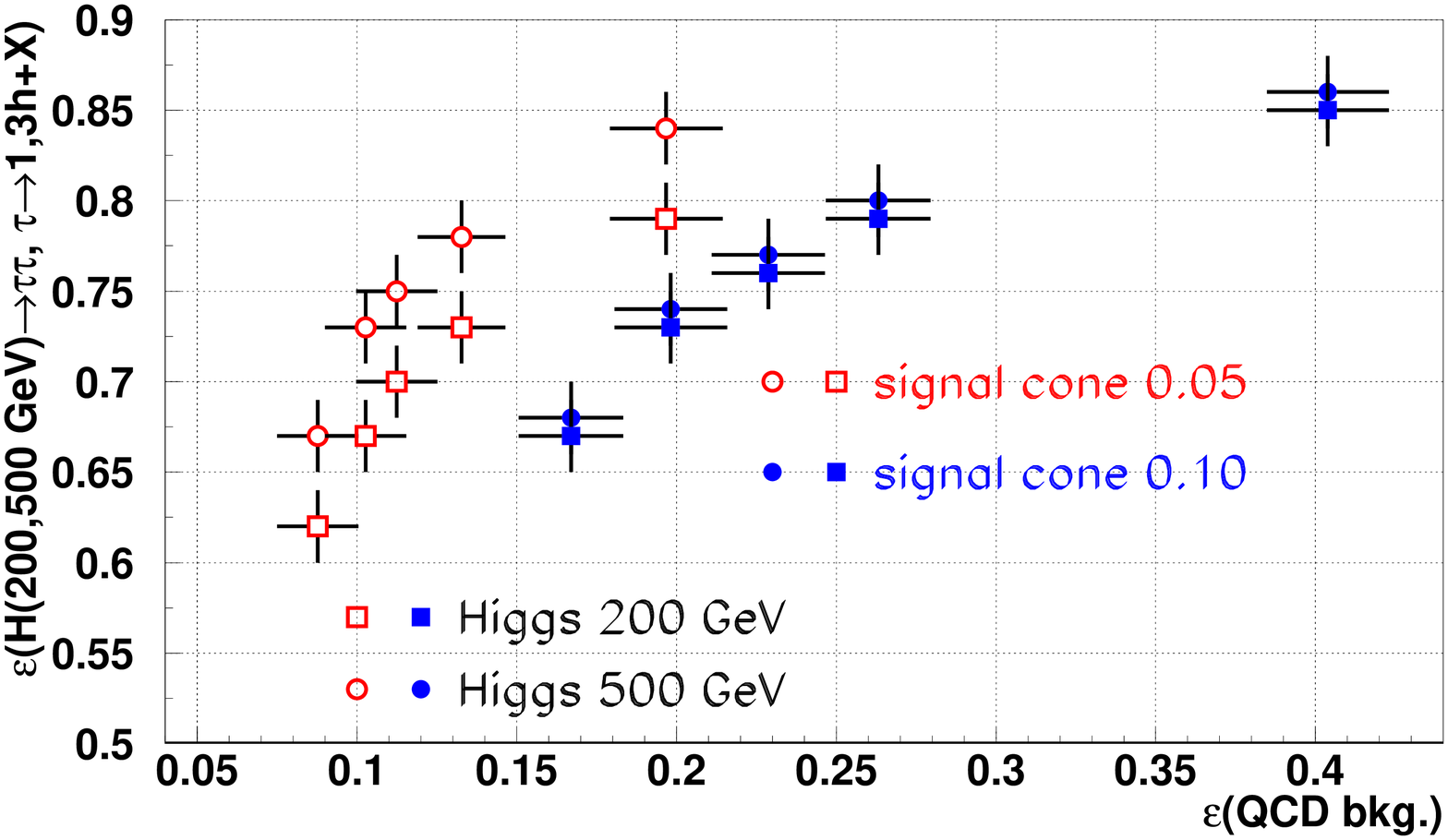}}
    \caption{The variation of the QCD background
             and the Higgs signal efficiency
             with the isolation cone size (0.20, 0.30, 0.35,
             0.40, 0.50) for Higgs mass $200 GeV/c^2$ (squares) and 
             $500 GeV/c^2$
             (circles) and for two values of the signal cone size 0.05
             (full marks) and 0.10 (open marks). Other parameters are
             $R_{m}$=0.10, $p^{m}_{t}=3 GeV/c$ and $p^{i}_{t}=1 GeV/c$.}
    \label{fig:lvl3_sig_bkg_eff}
  \end{center}
\end{figure}    
The variation of the QCD 2-jet
background and the Higgs signal efficiency with the change of the isolation
cone size is shown in Fig.~\ref{fig:lvl3_sig_bkg_eff} for Higgs mass of
$200 GeV/c^2$ (full squares) and $500 GeV/c^2$ 
(full circles) and for two signal cone
sizes $R_{s}$=0.05 and 0.10.
The points from left to right correspond to the following values of
$R_{i}$ = 0.50, 0.40, 0.35, 0.30 and 0.20.
One can see that a rejection factor of 5 (for the 0.10 signal cone) can be 
reached with an efficiency $\simeq$ 75\%. 
A higher background rejection of 10 (for the 0.05 signal cone) can be 
achieved but at a price of the efficiency being dependent on the Higgs mass.

\section{Conclusions}
\label{con}
Fast HLT algorithms for B physics decays and $\tau$-jets 
have been developed based on the CMS pixel detector and the innermost 
strip layer. The key feature of these algorithms is track reconstruction
based on 3 hits only. Track parameters are calculated based on a
simple helix approach. It is demonstrated that one can obtain high track 
reconstruction efficiency ($\sim$ 90\%), 
very accurate primary vertex position in the $z$ dimension 
($\le 50\mu m$) and a good momentum  resolution (1\%$\div$4\%) for 
tracks with  $p_t = 1 \div 10GeV/c$. 
High reduction of background rates (about 30 for B physics
channels and 5 for Higgs) with a good signal efficiency of 75\% are achieved.  

\ack      
{The $\tau$ trigger algorithm has been developed together with A. Nikitenko.
We would like to thank the organizers of the Vertex2001 conference
for the pleasant and fruitful meeting in Brunnen.}
\end{document}